\documentclass[conference]{IEEEtran}
\usepackage{cite}
\usepackage{amsmath,amssymb}
\usepackage{graphicx}
\usepackage{textcomp}
\usepackage[table]{xcolor}
\usepackage{booktabs}
\usepackage{tabularx}
\usepackage{orcidlink}
\usepackage{array}
\usepackage{url}
\usepackage{pifont}
\usepackage{enumitem}
\usepackage{tikz}
\usetikzlibrary{arrows.meta}
\usepackage[nointegrals]{wasysym}
\definecolor{trajcol}{HTML}{EEF6EF}
\definecolor{trajhd}{HTML}{CFE6D3}
\definecolor{trajmark}{HTML}{2E7D4F}
\newcommand{\full}{\textcolor{trajmark}{\large$\CIRCLE$}}
\newcommand{\part}{\textcolor{trajmark}{\large$\LEFTcircle$}}
\newcommand{\none}{\textcolor{trajmark}{\large$\Circle$}}
\definecolor{preai}{HTML}{F1F5FB}
\definecolor{preaihd}{HTML}{D3E0F0}
\definecolor{aiera}{HTML}{FDF4EC}
\definecolor{aierahd}{HTML}{F5DCC4}
\definecolor{grouprow}{HTML}{F2F2F2}
\definecolor{starcol}{HTML}{D98A1F}
\newcommand{\na}{\textcolor{black!30}{--}}
\newcommand{\hard}{\,\textcolor{starcol}{\ding{72}}}
\newcommand{\grp}[1]{\multicolumn{8}{>{\columncolor{grouprow}[0pt][0pt]}l@{}}{\rule{0pt}{2.2ex}\textbf{\textcolor{black!75}{#1}}}\\}

\begin{document}

\title{Information Needs in Code Review: What Can an Agent’s Trajectory Answer?
}

\author{
\IEEEauthorblockN{Hannah Estes~\orcidlink{0009-0006-9893-286X}, Saba Khanmohammadi~\orcidlink{0009-0003-5020-1695}, Amirali Rayegan~\orcidlink{0009-0004-8483-9934}, Rujuta Palimkar~\orcidlink{0009-0005-5565-6001}, Wesley Assunção~\orcidlink{0000-0002-7557-9091}}
\IEEEauthorblockA{\textit{Department of Computer Science} \\
\textit{North Carolina State University}\\
Raleigh, NC, United States}
\{hmestes, skhanmo, arayega, rspalimk, wguezas\}@ncsu.edu}

\maketitle
\bstctlcite{BSTcontrol}
\begin{abstract}
Coding agents open pull requests (PRs) at scale, and many are merged with minimal human review, yet people remain accountable for what gets merged. When review is skipped or left to agents, teams lose what PRs provide beyond defect detection: understanding, rationale, and knowledge transfer.
Behind each agent-authored PR lies an execution trajectory recording what the agent read, changed, ran, and abandoned. Conventional review surfaces only the diff and the agent’s own description. While raw trajectories are often too long for practical use during code review, they should be converted into context to support human readability and engagement. By mapping reviewer information needs from prior code-review literature to identify \textit{what} information a trajectory can supply and a diff alone cannot. 
Using this mapping, we propose a method to create short, extracted summaries in which each item is traceable to its corresponding trajectory steps. We adapt CAFE(S), a framework for assessing the context provided to coding agents (i.e., clarity, actionability, fidelity, efficiency, and security), to the summaries extracted for reviewers, human or not. We argue that such evidence must distinguish between what the agent claims and what its recorded tool outputs show, including whether each check was part of the project’s own tests or a script the agent wrote, since passing checks alone does not establish correctness. Lastly, a research agenda is presented to test this context with human reviewers and agentic reviewers.
\end{abstract}


\section{Introduction}
Software engineering agents explore repositories, edit files, run tests, and open pull requests (PRs) with minimal human involvement~\cite{wang2025openhands, li2026aidev}. Yet human code review has not scaled with code generation: most agent-authored PRs receive no review, and those that are reviewed are often reviewed only by other agents~\cite{duma2026}. For such reviews, a conventional agent-authored PR mostly includes the resulting diff and an agent-generated description~\cite{gong2026mci}; however, it often omits the execution history for that change. Reviewers can therefore only inspect \emph{what} changed without readily seeing \emph{what was inspected}, \emph{which alternatives failed}, or \emph{how claims for verification were supported}. Reviewers often reject these PRs that fail continuous integration, are unwanted, or are too large~\cite{ehsani2026fail,nakashima2026rejected}.

When comparing review behaviors in an analysis of 278{,}790 review conversations, more than 95\% of AI agent comments addressed code improvement or defect detection~\cite{zhong2026synergy}. Human reviewers, in contrast, also contributed to understanding, testing, and knowledge transfer~\cite{treude2026will, bacchelli2013expectations}. Across 1.02 million reviewed PRs, agent-involved reviews reached decisions faster under some adoption practices, but these efficiency gains did not translate into better review quality~\cite{zhong2026eras}.

When humans review agent-authored PRs, they face two specific problems. First, reviewing agent-generated changes that are often verbose without access to their underlying context contributes to cognitive and intent debt, which is the erosion of the team's shared understanding and coordination that threatens software health~\cite{storey2026technical}.
Second, an agent can hallucinate~\cite{tang2026coding}. For example, in 23,247 agent-authored PRs, the most common inconsistency was a \emph{phantom change}, a description claiming work absent from the diff, and these highly inconsistent PRs were far less likely to be merged~\cite{gong2026mci}. 
 
Code review has also never been only about finding defects. Reviewers seek to understand a change, its rationale, and its fit within the code repository~\cite{bacchelli2013expectations}. Pascarella et al.~\cite{pascarella2018information} identified seven such \emph{information needs}: (i) the suitability of alternative solutions, (ii) correct understanding, (iii) rationale, (iv) code context, (v) necessity, (vi) specialized expertise, and (vii) whether the change should be split. In human-authored PRs, many of these needs are met by asking the author or company-specific practices~\cite{pascarella2018information}. 
Agentic development introduces an additional source of review information: the agent’s execution trajectory. A trajectory records a sequence of thought–action–result steps in which the agent produces reasoning, invokes tools, and observes their outputs~\cite{bouzenia2025trajectories}. 
 
We frame this problem of identifying \textit{what} is the necessary information for code review through CAFE(S)~\cite{houck2026cafes}, a framework for evaluating the context provided to AI coding agents across five properties: (i) clarity, (ii) actionability, (iii) fidelity, (iv) efficiency, and (v) security. While CAFE(S) was designed for context flowing \emph{into} an agent, we apply it to context flowing \emph{out of} one, to the context needed for the (human or agentic) reviewer who must judge the result. In this sense, a review summary should be \emph{clear} (i.e., readable by a person without consulting the raw trajectory log), \emph{actionable} (i.e., organized around the questions a reviewer must answer to approve or reject), \emph{faithful} (i.e., every item is traceable to the trajectory step that produced it and the agent's own reasoning labeled as its explanation), \emph{efficient} (i.e., small enough to read alongside the diff), and \emph{secure} (i.e., exposing only trajectory content that is appropriate to share on a PR).
 
Rather than replacing PRs~\cite{sobo2026delta}, we argue for keeping the PR and supporting the many roles it serves, such as providing rationale and accountability for proposed changes. We call this approach \emph{trajectory-enriched code review}. To ground this idea, this paper reports our work organized around two questions:
 
\begin{itemize}
    \item[\textbf{RQ1}] \textit{Which reviewer information needs can a trajectory answer that the typical PR cannot?} We map reviewers’ information needs to evidence available in agent trajectories. We draw on Pascarella et al.’s seven needs~\cite{pascarella2018information} and add four that agent-authored PRs make pressing: (i) verification evidence, (ii) impact and completeness, (iii) description-code consistency, and (iv) provenance.
    \item[\textbf{RQ2}] \textit{How can that information be extracted faithfully and concisely?
    } Guided by the CAFE(S) properties, we describe an extraction process that cites the trajectory step for every item it reports.
\end{itemize}

\section{Related Work}

\paragraph{Reviewer Information Needs}
Code review serves goals beyond defect detection, including understanding a change, transferring knowledge, and assessing alternatives~\cite{bacchelli2013expectations}. Pascarella et al.\ derived seven information needs from 900 review threads~\cite{pascarella2018information}. Earlier industrial work found rationale to be the most important need, and consistency with the rest of the codebase among the hardest to obtain~\cite{tao2012understand}. Reviewers build a mental model of a change from the PR description, the issue tracker, continuous-integration results, and the surrounding code~\cite{wurzelgoncalves2025crcm}. These studies assume a human author whose intent can be recovered from the description and the review discussion. We test which of their needs an agent's trajectory can answer.

\paragraph{Oversight of AI-Authored Changes}
Labeling code as LLM-generated leads reviewers to spend longer on it, attend to its logical correctness, and use the generation prompt as a requirement or as documentation~\cite{khojah2026same}. Awareness of provenance also changes how developers validate code~\cite{tang2024developer}. In repositories, oversight of agent-authored PRs is limited: most receive no review, and the reviews that do occur are often performed by other agents~\cite{duma2026}. A study shows about 80\% of AI-co-authored PRs from non-owners merge without explicit review~\cite{gao2026autopilot}. Reviewer comments on agent-authored PRs concentrate on testing, security, and build concerns~\cite{haider2026dominant}. Descriptions that claim absent changes are the most frequent message, and these code inconsistencies are accepted less often~\cite{gong2026mci}. Failed agent-authored PRs more often fail continuous integration~\cite{ehsani2026fail}, and rejections cite PRs that are unwanted or too large~\cite{ehsani2026fail,nakashima2026rejected}. Developers who use agents report that agent output is hard to review and rely on heuristics such as trusting passing tests~\cite{dhanorkar2026oversight}. These studies document the oversight problem; they do not test what information would help a reviewer address it.

\paragraph{Agent Trajectories as Evidence}
Bouzenia and Pradel describe the actions in agent trajectories (i.e., exploring, locating, generating fixes) and recurring anti-patterns~\cite{bouzenia2025trajectories}. Gathering context before editing and validating changes were associated with success, while the relationship between trajectory length and failure reversed after accounting for task difficulty~\cite{mehtiyev2026beyond}. Trajectory-aware verifiers improve patch selection but can be misled by the agent's reasoning~\cite{jain2025r2egym}. RETRACE reconciles a forward reading of the trajectory with a backward reading of the patch to revise before submitting~\cite{li2026retrace}. These uses keep the trajectory inside the agent's own loop or treat behavior as a population-level correlate. We instead ask whether instance-level trajectory evidence is informative to a separate reviewer.

\paragraph{Patch Verification and LLM Reviewers}
Passing a benchmark's tests does not guarantee a patch is correct~\cite{wang2026solved}. Practitioners report that missing context about why a change was made limits LLM-assisted review~\cite{adalsteinsson2025rethinking}. SWE-Review trains agentic reviewers that explore the repository and outperform fixed-context review~\cite{wang2026swereview}. The LLM reviewers' decisions can shift with the stated approval policy or the wording of the review prompt~\cite{kudum2026policies,jin2026overcorrection}. Guo et al.\ found that the full trajectory did not yield the best trade-off between catching incorrect patches and rejecting correct ones~\cite{guo2026groundability}. Kamal{\i} et al.\ use LLM agents that re-run a PR's code to collect logs and traces and summarize them as verifiable claims~\cite{kamali2026rethinking}. We differ in two ways: (i) we derive the evidence from reviewer information needs rather than from an auditing design to support human reviewers, and (ii) we use the already created trajectory to extract the information rather than additional LLM-based processing.

\section{RQ1: Reviewer Needs and Trajectory Evidence}\label{sec:needs}
\paragraph{Method}
We adopt Pascarella et al.’s seven reviewer information needs (N1--N7) as the foundation of our mapping~\cite{pascarella2018information}. Additionally, we included four needs that recur in AI-based code-review studies, outlined in Table~\ref{tab:alignment}, but are missing from that set: verification evidence (A1), impact and completeness (A2), description--code consistency (A3), and AI provenance (A4). These added needs (i.e., A1-A4) required at least two sources to mention them, and at least one of which describes them as information a reviewer seeks or acts on during the review. We aligned each need with constructs from three pre-AI studies~\cite{tao2012understand,bacchelli2013expectations,wurzelgoncalves2025crcm} and three AI-era studies~\cite{khojah2026same,haider2026dominant,gong2026mci}, using each paper's own label, as shown in Table~\ref{tab:alignment}). We selected these six studies because they report reviewer information needs or related constructs at a level of granularity comparable to that of Pascarella et al.~\cite{pascarella2018information}. Together, they span several research methods, including surveys, interviews, observation, eye tracking, and repository mining.
Further described in Section~\ref{sec:answer}, we rate a need as answered \textit{directly} when the trajectory records the information the need asks for, \textit{partly} when it records signals a reviewer must interpret to address the need, and \textit{not at all} when it contains no relevant record. A direct rating means the evidence is present, not that the need is resolved: a recorded passing test still leaves the reviewer to judge whether the change was exercised.

\begin{table*}[t]
\centering
\scriptsize
\caption{Reviewer Information Needs and What a Trajectory Can Answer}
\label{tab:alignment}
\setlength{\tabcolsep}{3.2pt}
\renewcommand{\arraystretch}{1.00}
\begin{tabularx}{\textwidth}{@{}>{\raggedright\arraybackslash}p{1.8cm}
  *{3}{>{\columncolor{preai}\raggedright\arraybackslash}X}
  *{3}{>{\columncolor{aiera}\raggedright\arraybackslash}X}
  >{\columncolor{trajcol}[\tabcolsep][0pt]\centering\arraybackslash}p{1.15cm}@{}}
\toprule
\multicolumn{1}{@{}l}{} & \multicolumn{3}{c}{\cellcolor{preaihd}\rule[-0.8ex]{0pt}{3ex}\textbf{\textcolor[HTML]{1F3F63}{Pre-AI Code Review}}}
 & \multicolumn{3}{c}{\cellcolor{aierahd}\textbf{\textcolor[HTML]{7A3E0E}{AI / Agentic Era}}}
 & \multicolumn{1}{>{\columncolor{trajhd}[\tabcolsep][0pt]}c@{}}{\textbf{\textcolor[HTML]{1E5631}{This Paper}}}\\
\rowcolor{white}
\textbf{Needs} & \textbf{Tao et al.~\cite{tao2012understand}} & \textbf{Bacchelli \& Bird~\cite{bacchelli2013expectations}} & \textbf{Wurzel Gon\c{c}alves et al.~\cite{wurzelgoncalves2025crcm}} & \textbf{Khojah et al.~\cite{khojah2026same}} & \textbf{Haider \& Zimmermann~\cite{haider2026dominant}} & \textbf{Gong et al.~\cite{gong2026mci}} & \textbf{Trajectory answers}\\
\midrule
\grp{A. Understanding the change}
\textbf{N2} Correct understanding & Behavior (I-6) & Understanding & Mental model of the PR & Prompt as documentation & \na & \na & \part\\
\textbf{N3} Rationale & Rationale (I-1) & Track Rationale & PR title and description; issue tracking & Prompt (``original intent'') & \na & \na & \part\\
\textbf{N4} Code context & References (I-7); Caller (I-8) & Context and change understanding & Code base, tests, or documentation; other PRs & \na & \na & \na & \part\\
\grp{B. Judging the solution}
\textbf{N1} Suitability of an alternative solution & Ideal/optimal solution & Alternative Solutions & Ideal mental model & \na & \na & \na & \part\\
\textbf{N5} Necessity\hard & \na & Code Improvement & \na & \na & Code Restructure, Simplification \& Redundancy Elimination & \na & \part\\
\textbf{A1} Verification evidence\hard & Correctness (I-3); Tests (I-11); New tests (I-12); Failing tests (I-13) & Avoid Build Breaks; Testing & CI/CD status; Testing (II.T) & Logical correctness & Testing, Assertions \& Validation & File-type consistency & \part~\\
\textbf{A2} Impact and completeness & Risk (I-9); Completeness (I-2); Consistency (I-10) & External Impact & \na & \na & Security, Safety \& Reliability; Build, Configuration \& Default Behavior Mgmt.; Dependency, Module \& Import Mgmt. & \na & \part\\
\grp{C. Scope and provenance}
\textbf{N7} Splittable\hard & Change decomposition & \na & Review size; Chunking (II.CR.C) & \na & \na & \na & \part\\
\textbf{N6} Specialized expertise & \na & \na & Knowledge of their colleagues' expertise & \na & \na & \na & \none\\
\textbf{A3} Description--code consistency\hard & Quality of the change description & \na & Annotation layer & Prompt as requirement & \na & Phantom Changes; Scope Understated; Placeholder/\allowbreak{}Incomplete; four further types & \part~\\
\textbf{A4} AI provenance\hard & \na & \na & \na & LLM label; full conversation history & \na & High-MCI prevalence by agent & \full\\
\midrule
\textcolor{black!60}{\textit{Other concepts reported}} & \textcolor{black!60}{Design (I-4); Clones (I-5); Change-/Defect-proneness (I-14, I-15)} & \textcolor{black!60}{Finding defects, Knowledge Transfer, Team Awareness, Share Code Ownership, Improving Dev Process, Team Assessment} & \textcolor{black!60}{Code reading strategies; external sources} & \textcolor{black!60}{Code quality; readability; consistency with the rest of the codebase} & \textcolor{black!60}{Seven further themes (e.g., Styling \& Formatting Adjustments; Performance Improvement)} & \textcolor{black!60}{Other/Mixed} & \\
\bottomrule
\end{tabularx}
\par\smallskip
\parbox{\textwidth}{\scriptsize\raggedright
\textit{Note.} N1--N7 map to Pascarella et al.~\cite{pascarella2018information}; A1--A4 are added by us. Each cell shows the cited paper's closely related label; a dash indicates the study does not address the need. \hard~marks a need that AI-era studies report as harder to meet in agent-authored PRs (Section~\ref{sec:needs}). Last column: \full~answered directly, \part~partly, \none~not at all by trajectory evidence (Section~\ref{sec:answer}).
}
\end{table*}

\paragraph{Findings}
Pre-AI studies already cover the seven primary information needs.
AI-era studies show that five needs become harder to meet for agent-authored PRs (N5, N7, A1, A3, A4). Description--code consistency (A3) breaks down because agent descriptions claim changes that are absent from the code~\cite{gong2026mci}. For this, a label will be added to the extracted information to indicate whether a summary item was observed or whether the agent's rationale was inferred from the trajectory. Verification evidence (A1) becomes less trustworthy because test-passing patches can be incorrect~\cite{wang2026solved} and failed agent-authored PRs more often fail continuous integration~\cite{ehsani2026fail}. This matters more since reviewers of AI-labeled code primarily focus on logical correctness~\cite{khojah2026same}. Necessity (N5) and splittability (N7) gain importance because reviewers typically reject agent-authored PRs as unwanted or too large~\cite{ehsani2026fail,nakashima2026rejected}. A1--A3 are added relative to Pascarella et al.'s set but similar needs appear in pre-AI studies; only provenance (A4) has no pre-AI counterpart~\cite{khojah2026same,tang2024developer}. These five are also the needs an agent-written PR description is least able to answer, as it reports on its own work. The last row of Table~\ref{tab:alignment} lists constructs that match no need, which mostly revolve around code quality, code-reading strategies, and team-specific processes.

\section{RQ2: Extracting Trajectory Evidence}\label{sec:answer}

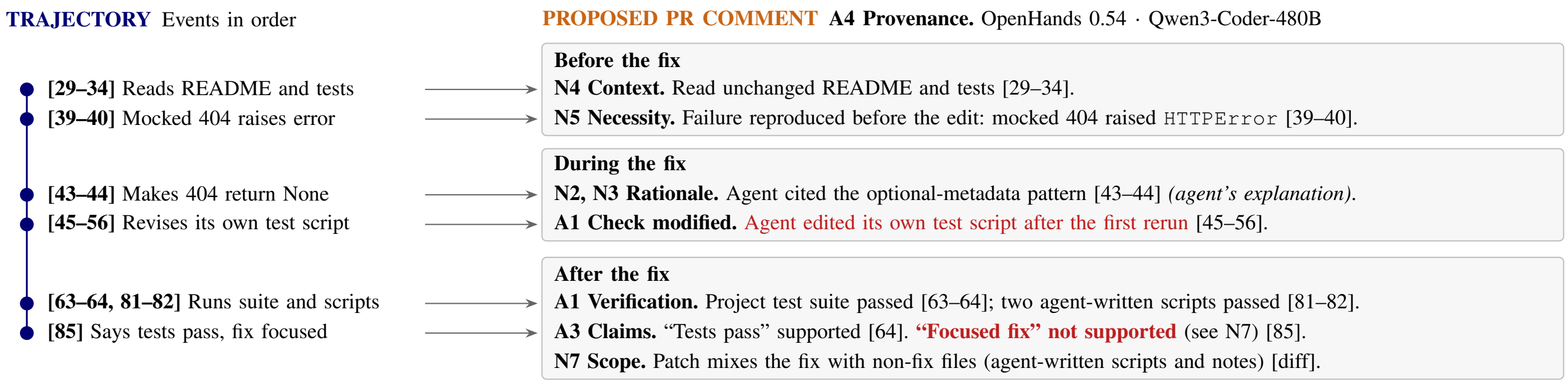
\begin{figure*}[t]
\begin{tikzpicture}[
  y=0.76cm,
  font=\scriptsize,
  >={Stealth[length=4pt]},
  ev/.style={anchor=west, inner sep=1pt, text width=4cm, align=left},
  item/.style={anchor=west, inner sep=1pt, text width=11cm, align=left},
  sec/.style={draw=black!25, fill=black!3, rounded corners=1.5pt},
  sechead/.style={anchor=west, inner sep=1pt, font=\scriptsize\bfseries},
  dot/.style={circle, fill=blue!45!black, inner sep=1.6pt},
]
\definecolor{flag}{RGB}{190,30,30}
\definecolor{hdr}{RGB}{200,100,20}

\node[anchor=west, inner sep=0] at (0,0)
  {\textbf{\textcolor{blue!45!black}{TRAJECTORY}}\enspace Events in order};
\node[anchor=west, inner sep=0] at (6.2,0)
  {\textbf{\textcolor{hdr}{PROPOSED PR COMMENT}}\enspace
   \textbf{A4 Provenance.} OpenHands 0.54 $\cdot$ Qwen3-Coder-480B};

\draw[sec] (6.2,-0.35) rectangle (18.0,-1.75);
\draw[sec] (6.2,-1.95) rectangle (18.0,-3.35);
\draw[sec] (6.2,-3.6) rectangle (18.0,-5.45);
\node[sechead] at (6.3,-0.6)  {Before the fix};
\node[sechead] at (6.3,-2.2)  {During the fix};
\node[sechead] at (6.3,-3.85)  {After the fix};

\draw[blue!45!black, line width=0.6pt] (0.25,-1.05) -- (0.25,-4.75);

\newcommand{\event}[3]{%
  \node[dot] at (0.25,#1) {};
  \node[ev] at (0.45,#1) {\textbf{[#2]} #3};
  \draw[->, black!60] (4.85,#1) -- (6.15,#1);}

\event{-1.05}{29--34}{Reads README and tests}
\event{-1.50}{39--40}{Mocked 404 raises error}
\event{-2.65}{43--44}{Makes 404 return None}
\event{-3.10}{45--56}{Revises its own test script}
\event{-4.30}{63--64, 81--82}{Runs suite and scripts}
\event{-4.75}{85}{Says tests pass, fix focused}

\node[item] at (6.3,-1.05) {\textbf{N4 Context.} Read unchanged README and tests [29--34].};
\node[item] at (6.3,-1.50) {\textbf{N5 Necessity.} Failure reproduced before the edit: mocked 404 raised \texttt{HTTPError} [39--40].};

\node[item] at (6.3,-2.65) {\textbf{N2, N3 Rationale.} Agent cited the optional-metadata pattern [43--44] \textit{(agent's explanation)}.};
\node[item] at (6.3,-3.10) {\textbf{A1 Check modified.} \textcolor{flag}{Agent edited its own test script after the first rerun} [45--56].};

\node[item] at (6.3,-4.30) {\textbf{A1 Verification.} Project test suite passed [63--64]; two agent-written scripts passed [81--82].};
\node[item] at (6.3,-4.75) {\textbf{A3 Claims.} ``Tests pass'' supported [64]. \textcolor{flag}{\textbf{``Focused fix'' not supported}} (see N7) [85].};
\node[item] at (6.3,-5.20) {\textbf{N7 Scope.} Patch mixes the fix with non-fix files (agent-written scripts and notes) [diff].};
\end{tikzpicture}
\caption{From agent trajectory to review evidence. Selected events from the SWE-rebench~\cite{badertdinov2025swerebench} OpenHands run for \texttt{adamchainz/ec2-metadata} issue \#350 are mapped to a proposed PR comment. Bracketed numbers identify messages in the original trajectory.}
\label{fig:trajectory-extraction}
\end{figure*}

We map each need in Table~\ref{tab:alignment} to the trajectory actions identified by Bouzenia and Pradel~\cite{bouzenia2025trajectories} and to evidence that deterministic rules can extract. A trajectory answers provenance directly; verification, description--code consistency, rationale, context, necessity, alternatives, impact, and splittability partly; and specialized expertise not at all.

Our proposed extractor parses the agent's message and tool-call history, maintains an index for each step, and classifies file views, searches, shell commands, edits, reverts, test runs, and the completion message, recording exit codes and failing tests from tool outputs. The verifier confirms that each cited step exists and contains the quoted content; it may not verify claims that were drawn from each step. Rules over these events produce evidence for each need:

\textbf{N1 Alternative solutions:} discarded or reverted edits and the checks that followed them. Draws on the \emph{Divergence} relationship and repeated \emph{Generate fix} actions.
\textbf{N2, N3 Understanding and Rationale:} the agent's stated reasoning before each retained edit (i.e., the agent's explanation). Draws on the \emph{Explain} action and the \emph{Thought} component.
\textbf{N4 Code context:} files read but not changed, and search hits never opened. Draws on the \emph{Explore}, \emph{Locate}, and \emph{Search} actions.
\textbf{N5 Necessity:} whether a failure was observed before the first edit. Draws on the \emph{Reproduce} action.
\textbf{N6 Specialized expertise:} Not answerable from an agent's trajectory.
\textbf{A1 Verification evidence:} the checks run after the last edit, what each one ran (a script the agent wrote or the project's own tests), who wrote it, its outcome, and whether the agent modified a check between runs. Draws on the \emph{Reproduce} and \emph{Run tests} actions and on termination without proper test validation.
\textbf{A2 Impact and completeness:} for each edited symbol, whether the agent searched for its references, how many hits it opened, and whether any check it ran exercised those callers. Draws on the \emph{Locate} and \emph{Search} actions. Completeness: issue requirements the agent never acted on, and runs that ended by timeout or step limit.
\textbf{N7 Splittability:} edits grouped by the sub-goal they served, where each group has its own reproduce–edit–verify cycle, and files added that serve no sub-goal (e.g., agent scripts, notes). Draws on repeated \emph{Generate fix} cycles.
\textbf{A3 Description--code consistency:} claims in the agent's final message, checked against the recorded events. Draws on the \emph{Misalignment} (thought--action) and \emph{Misinterpretation} (result--thought) relationships. If the extractor is deterministic, this will be limited to structured events (e.g., ``test pass'' against test-run events, ``focused-fix'' against the diff's file list). Alternatively, a hybrid approach would include a small model for extracting specific claims, and then a subsequent check based on specific trajectory steps.
\textbf{A4 AI provenance:} the agent, model and task prompt, taken from trajectory metadata.


To motivate our design of the extracted summaries, we adapt CAFE(S)~\cite{houck2026cafes}, which assesses the context given \emph{to} coding agents, into requirements for the context that flows \emph{out of} an agent to its reviewer. Review evidence should be \emph{clear}: readable without consulting the raw log. It should be \emph{actionable}: organized around the needs above and omitting what the diff already shows, such as the list of changed files. It should be \emph{faithful}: every item cites the trajectory steps it came from, a checker confirms that each cited step exists and contains the quoted text, command or path, and the agent's own reasoning is labeled as its explanation, not as observed fact. It should be \emph{efficient}: small enough to read beside the diff. Finally, it should be \emph{secure}: exposing only trajectory content that is appropriate to post on a PR, which requires redaction (e.g., secrets, internal URLs) that citations alone do not provide. Figure~\ref{fig:trajectory-extraction} shows evidence for one agent-authored PR built to these requirements.

\section{Future Work}
Our position raises a question that the mapping and extracted summaries alone cannot answer: does trajectory evidence actually change how agent-authored PRs are reviewed, and for whom? We plan to address this through three future questions (FQ1--FQ3).

\textbf{FQ1: Does trajectory evidence help human reviewers oversee agent-authored PRs?}
Reviewers of real agent-authored PRs will see the diff alone, the raw trajectory, or our extracted evidence. Accuracy alone can hide the failure we care most about, so our primary measure is how often reviewers approve incorrect changes, alongside comprehension, review time, and whether reviewers' confidence matches the correctness of their decisions. The goal is to calibrate trust in the agent's work by providing evidence that separates what the agent claims from what the checks show, so reviewers know when to rely on the agent's work and when to question it.

\textbf{FQ2: Does the same evidence serve agent reviewers, either as a stand-in for or a complement to human review?}
CAFE(S)~\cite{houck2026cafes} was built to assess context given to coding agents, which suggests that the extracted evidence shaped by this framework should be usable by agents as well. To test this, we will give several LLM reviewers the same setup as FQ1. Agreement with the human decisions from FQ1 will indicate which support agent reviewers prefer or are better suited to assisting them. The cost per review, depending on the model, will show whether this is practical at scale.

\textbf{FQ3: Do individual differences among reviewers change how they use trajectory evidence?}
Not every reviewer reads evidence the same way. Drawing on GenderMag's cognitive facets~\cite{burnett2016gendermag}, we expect reviewers who differ in self-efficacy and attitudes toward risk to rely differently on an agent's claim that its tests passed, and reviewers with different information processing styles to prefer a short summary or a complete list of steps. Reusing the setup from FQ1, we will compare our default presentation with facet-informed variants to see whether presentation can help across reviewers.

Beyond these questions, trajectories may contain useful signals outside the needs we mapped, and analyzing them could extend the framework beyond its current categories.

\section{Conclusion}
Coding agents are producing PRs faster than people can meaningfully review them, yet people remain responsible for what gets merged. In this paper, we argued that agent-authored PRs should keep the PR but carry evidence extracted from the agent’s trajectory, organized around what reviewers need to know. By mapping reviewer information needs from prior code-review literature, we identified which of these needs a trajectory can answer and which become harder to meet for agent-authored PRs. 
We adapted CAFE(S), a framework originally designed to assess the context provided to coding agents, to describe what good review evidence should look like to serve both human and agent reviewers.
Such evidence should state what each check ran and whether the agent wrote it, not only that it passed. Our planned studies will examine whether this evidence helps human reviewers, whether agent reviewers can use it, and how its presentation can support reviewers who approach code review differently.

\bibliographystyle{IEEEtran}
\bibliography{refs}

\end{document}